\documentclass[aps,prl,twocolumn,linenumber,floatfix,citeautoscript,superscriptaddress]{revtex4-1}
\usepackage{graphicx}
\usepackage{bm,amsmath,amssymb,mathrsfs,dcolumn}
\usepackage[colorlinks=true, linkcolor=blue, citecolor=blue, urlcolor=blue, linktoc=page, bookmarks=false, pdfstartview={FitH}, pdfborder={0 0 0.0 [3 3]}]{hyperref} 
\usepackage[usenames]{color}
\usepackage{epstopdf}
\usepackage{subfigure}
\usepackage{units}
\usepackage{lineno}
\usepackage{url}
\usepackage{multirow}
\usepackage{graphicx}
\usepackage{booktabs}

\begin{document}

\title{First measurement of the yield of $^8$He isotopes produced in liquid scintillator by cosmic-ray muons at Daya Bay}
\newcommand{\IHEP}{\affiliation{Institute~of~High~Energy~Physics, Beijing}}
\newcommand{\Wisconsin}{\affiliation{University~of~Wisconsin, Madison, Wisconsin 53706}}
\newcommand{\Yale}{\affiliation{Wright~Laboratory and Department~of~Physics, Yale~University, New~Haven, Connecticut 06520}} 
\newcommand{\BNL}{\affiliation{Brookhaven~National~Laboratory, Upton, New York 11973}}
\newcommand{\NTU}{\affiliation{Department of Physics, National~Taiwan~University, Taipei}}
\newcommand{\NUU}{\affiliation{National~United~University, Miao-Li}}
\newcommand{\Dubna}{\affiliation{Joint~Institute~for~Nuclear~Research, Dubna, Moscow~Region}}
\newcommand{\CalTech}{\affiliation{California~Institute~of~Technology, Pasadena, California 91125}}
\newcommand{\CUHK}{\affiliation{Chinese~University~of~Hong~Kong, Hong~Kong}}
\newcommand{\NCTU}{\affiliation{Institute~of~Physics, National~Chiao-Tung~University, Hsinchu}}
\newcommand{\NJU}{\affiliation{Nanjing~University, Nanjing}}
\newcommand{\TsingHua}{\affiliation{Department~of~Engineering~Physics, Tsinghua~University, Beijing}}
\newcommand{\SZU}{\affiliation{Shenzhen~University, Shenzhen}}
\newcommand{\NCEPU}{\affiliation{North~China~Electric~Power~University, Beijing}}
\newcommand{\Siena}{\affiliation{Siena~College, Loudonville, New York  12211}}
\newcommand{\IIT}{\affiliation{Department of Physics, Illinois~Institute~of~Technology, Chicago, Illinois  60616}}
\newcommand{\LBNL}{\affiliation{Lawrence~Berkeley~National~Laboratory, Berkeley, California 94720}}
\newcommand{\UIUC}{\affiliation{Department of Physics, University~of~Illinois~at~Urbana-Champaign, Urbana, Illinois 61801}}
\newcommand{\SJTU}{\affiliation{Department of Physics and Astronomy, Shanghai Jiao Tong University, Shanghai Laboratory for Particle Physics and Cosmology, Shanghai}}
\newcommand{\BNU}{\affiliation{Beijing~Normal~University, Beijing}}
\newcommand{\WM}{\affiliation{College~of~William~and~Mary, Williamsburg, Virginia  23187}}
\newcommand{\Princeton}{\affiliation{Joseph Henry Laboratories, Princeton~University, Princeton, New~Jersey 08544}}
\newcommand{\VirginiaTech}{\affiliation{Center for Neutrino Physics, Virginia~Tech, Blacksburg, Virginia  24061}}
\newcommand{\CIAE}{\affiliation{China~Institute~of~Atomic~Energy, Beijing}}
\newcommand{\SDU}{\affiliation{Shandong~University, Jinan}}
\newcommand{\NanKai}{\affiliation{School of Physics, Nankai~University, Tianjin}}
\newcommand{\UC}{\affiliation{Department of Physics, University~of~Cincinnati, Cincinnati, Ohio 45221}}
\newcommand{\DGUT}{\affiliation{Dongguan~University~of~Technology, Dongguan}}
\newcommand{\XJTU}{\affiliation{Department of Nuclear Science and Technology, School of Energy and Power Engineering, Xi'an Jiaotong University, Xi'an}}
\newcommand{\UCB}{\affiliation{Department of Physics, University~of~California, Berkeley, California  94720}}
\newcommand{\HKU}{\affiliation{Department of Physics, The~University~of~Hong~Kong, Pokfulam, Hong~Kong}}
\newcommand{\Charles}{\affiliation{Charles~University, Faculty~of~Mathematics~and~Physics, Prague}} 
\newcommand{\USTC}{\affiliation{University~of~Science~and~Technology~of~China, Hefei}}
\newcommand{\TempleUniversity}{\affiliation{Department~of~Physics, College~of~Science~and~Technology, Temple~University, Philadelphia, Pennsylvania  19122}}
\newcommand{\CGNPG}{\affiliation{China General Nuclear Power Group, Shenzhen}}
\newcommand{\NUDT}{\affiliation{College of Electronic Science and Engineering, National University of Defense Technology, Changsha}} 
\newcommand{\IowaState}{\affiliation{Iowa~State~University, Ames, Iowa  50011}}
\newcommand{\ZSU}{\affiliation{Sun Yat-Sen (Zhongshan) University, Guangzhou}}
\newcommand{\CQU}{\affiliation{Chongqing University, Chongqing}} 
\newcommand{\BCC}{\altaffiliation[Now at ]{Department of Chemistry and Chemical Technology, Bronx Community College, Bronx, New York  10453}} 

\newcommand{\UCI}{\affiliation{Department of Physics and Astronomy, University of California, Irvine, California 92697}} 
\newcommand{\GXU}{\affiliation{Guangxi University, No.100 Daxue East Road, Nanning}} 
\newcommand{\HKUST}{\affiliation{The Hong Kong University of Science and Technology, Clear Water Bay, Hong Kong}} 
\newcommand{\Rochester}{\altaffiliation[Now at ]{Department of Physics and Astronomy, University of Rochester, Rochester, New York 14627}} 

\newcommand{\LSU}{\altaffiliation[Now at ]{Department of Physics and Astronomy, Louisiana State University, Baton Rouge, LA 70803}} 

\author{F.~P.~An}\ZSU
\author{W.~D.~Bai}\ZSU
\author{A.~B.~Balantekin}\Wisconsin
\author{M.~Bishai}\BNL
\author{S.~Blyth}\NTU
\author{G.~F.~Cao}\IHEP
\author{J.~Cao}\IHEP
\author{J.~F.~Chang}\IHEP
\author{Y.~Chang}\NUU
\author{H.~S.~Chen}\IHEP
% ...

\date{\today }

\begin{abstract}

Daya Bay presents the first measurement of cosmogenic $^8$He isotope production in liquid scintillator, using an innovative method for identifying cascade decays of $^8$He and its child isotope, $^8$Li.
We also measure the production yield of $^9$Li isotopes using well-established methodology. The results, in units of 10$^{-8}\mu^{-1}$g$^{-1}$cm$^{2}$, are 0.307$\pm$0.042, 0.341$\pm$0.040, and 0.546$\pm$0.076 for $^8$He, and 6.73$\pm$0.73, 6.75$\pm$0.70, and 13.74$\pm$0.82 for $^9$Li at average muon energies of 63.9~GeV, 64.7~GeV, and 143.0~GeV, respectively.
The measured production rate of $^8$He isotopes is more than an order of magnitude lower than any other measurement of cosmogenic isotope production. 
It replaces the results of previous attempts to determine the ratio of $^8$He to $^9$Li production that yielded a wide range of limits from 0 to 30\%.
The results provide future liquid-scintillator-based experiments with improved ability to predict cosmogenic backgrounds. 

\end{abstract}


\maketitle


%
Unstable light isotopes, $^8$He and $^9$Li, produced in liquid scintillator by cosmic-ray muons, can undergo $\beta$ decay with associated neutron production.
About 16\% of $^8$He and 51\% of $^9$Li isotopes decay via the $\beta$-n channel~(Table~\ref{table:he8decay}).
The energy deposit of the $\beta$~(and $\alpha$'s for $^9$Li) and the subsequent capture of the neutron mimics the prompt and delayed signal, respectively, from inverse beta decay~(IBD) of electron antineutrinos~($\overline{\nu}_{e}$). 
This creates an irreducible background to the detection of $\overline{\nu}_{e}$ produced in reactors that are used to make precision measurements of neutrino oscillation~\cite{2012DYB,2012Chooz,2012RENO,KamLAND:2008dgz,JUNO:2022hxd}.

Due to the challenges introduced by these isotopes, measurements of their production yields with respect to muon energies have been performed in many neutrino experiments and accelerator beams~\cite{Hagner:2000xb,KamLAND:2009zwo,Borexino:2013cke,DoubleChooz:2018kvj,RENO:2022xbr,KamLAND-Zen:2023spw,Super-Kamiokande:2015xra}.
While the yield of the sum of $^9$Li and $^8$He production has been measured to a precision of 10\% or better, there has been no measurement of non-zero cosmogenic $^8$He production.
Previous investigations~\cite{KamLAND:2009zwo,Borexino:2013cke,DoubleChooz:2018kvj,RENO:2022xbr,KamLAND-Zen:2023spw} all used the $\beta$-n channel and had strong anti-correlation of the measured yields of $^8$He and $^9$Li.
The measured ratios of the yield of $^8$He relative to $^9$Li were consistent with zero at three standard deviations or less, but could be as large as 20\%~\cite{RENO:2022xbr} or 30\%~\cite{KamLAND:2009zwo,KamLAND-Zen:2023spw}. 
The ratios from the widely used GEANT4~\cite{Allison:2016lfl} simulation vary by a factor of a few depending upon the hadronic interaction model.

A innovative method was developed using the cascade decays of $^8$He and its child $^8$Li~(Channel $\bf A$ in Table~\ref{table:he8decay}).
$^8$He has a 171.7~ms lifetime with an 84\% branching ratio via $\beta$ decay to the first excited state of $^8$Li with an end-point energy of 9.68~MeV.
The first excited state of $^8$Li promptly releases a $\gamma$ particle with an energy of 0.98~MeV to the ground state of $^8$Li.
Subsequently $^8$Li $\beta$ decays with a lifetime of 1.2~s and an end-point energy of 12.97~MeV.
The cascade $\beta$ decays of $^8$He and $^8$Li provide a distinctive correlation in both space and time that allows discrimination of the small signal from background.
The high endpoint energies of the beta spectra enables discrimination from low energy background due to natural radioactive decays. 

\begin{table}[hbt]
\renewcommand\arraystretch{1.25}
\caption{Branching ratios~(BR) and decay schemes of $^8$He and $^9$Li with the detectable prompt and delayed products. Decay information is taken from the NuDat database~\cite{nuDAT}. Previous measurements~\cite{KamLAND:2009zwo,Borexino:2013cke,DoubleChooz:2018kvj,RENO:2022xbr,KamLAND-Zen:2023spw} relied on channels $\bf B$ and $\bf C$. The measurement in this study uses channel $\bf A$. There are multiple decay paths represented by $^{9}$Be$^{*}$ $\rightarrow$ $n\alpha\alpha$ for channel $\bf C$.}
\centering
\begin{tabular}{ccccc}
 & BR & Decay chain & Prompt & Delayed \\
\hline
\multirow{2}{*}{$\bf A$} & \multirow{2}{*}{84\%} & $^8$He $\rightarrow$ $^8$Li $\beta^{-}_{1}$ $\overline{\nu}$, $\gamma$ & \multirow{2}{*}{$\beta^{-}_{1}$, $\gamma$} & \multirow{2}{*}{$\beta^{-}_{2},\alpha,\alpha$} \\
 & & $^8$Li $\rightarrow$ $\alpha\alpha$ $\beta^{-}_{2}$ $\overline{\nu}$ & & \\
\hline
\multirow{2}{*}{$\bf B$} & \multirow{2}{*}{16\%} & $^8$He $\rightarrow$ $^8$Li$^*$ $\beta^{-}$ $\overline{\nu}$ & \multirow{2}{*}{$\beta^{-}$} & \multirow{2}{*}{$n$} \\
 & & $^8$Li$^*$ $\rightarrow$ $^7$Li $n$ & & \\
\hline
\multirow{2}{*}{$\bf C$} & \multirow{2}{*}{50.8\%} & $^9$Li $\rightarrow$ $^9$Be$^*$ $\beta^{-}$ $\overline{\nu}$ & \multirow{2}{*}{$\beta^{-},\alpha,\alpha$} & \multirow{2}{*}{$n$} \\
 & & $^9$Be$^{*}$ $\rightarrow$ $n\alpha\alpha$ & & \\
\hline
\multirow{2}{*}{$\bf D$} & \multirow{2}{*}{49.2\%} & $^9$Li $\rightarrow$ $^9$Be $\beta^{-}$ $\overline{\nu}$ & \multirow{2}{*}{$\beta^{-}$} & \multirow{2}{*}{none} \\
 & & $^9$Be stable & & \\

\hline
\end{tabular}
\label{table:he8decay}
\end{table}

The Daya Bay reactor neutrino experiment employed eight identically designed antineutrino detectors~(ADs) in three experimental halls~(EHs)~\cite{2016DYB:Detector}.
The three EHs were at vertical depths of 250, 265, and 860 meters-water-equivalent~(m.w.e.) allowing for a measurement of isotope yields at three different values of average muon energy within the same experiment.
Underground muon fluxes and energy spectra were simulated using a modified version of Gaisser’s formula to take into account the muon spectrum at low energies and large zenith angles~\cite{Guan:2015vja}.
MUSIC~\cite{KUDRYAVTSEV2009339} was used to propagate muons from the surface to the underground halls.
The average energies of muons reaching ADs ($\rm E^\mu_{\rm avg}$) in three EHs are (63.9$\pm$3.8)~GeV, (64.7$\pm$3.9)~GeV, and (143.0$\pm$8.6)~GeV, respectively.
Details of the muon simulation can be found in Ref.~\cite{DayaBay:2017txw}.

Each AD was a stainless steel vessel containing three concentric cylindrical regions separated by transparent acrylic vessels, filled with 20~tons of liquid scintillator loaded with 0.1\% gadolinium by weight~(GdLS), 22~tons of liquid scintillator~(LS), and mineral oil.
Each AD was immersed in a water shield~\cite{2015DYB:muonSystem}, which also serves as a water Cherenkov counter providing full coverage for muons traversing any AD.
When the zenith and azimuth angle distributions of the incident muons are taken into account, the average track lengths of muons passing through GdLS~($L^\mu_{\rm GdLS}$) are 204.1~cm, 204.5~cm, and 204.9~cm in three EHs.
Adding the traversal of the LS, the average track lengths~($L^\mu_{\rm LS}$) are 263.6~cm, 264.2~cm, and 264.4~cm.

The measurement of $^8$He production used the full Daya Bay data set acquired in 3,158 days of operation~\cite{DayaBay:2023orm}.
Events in an AD with reconstructed energy between 20~MeV and 1~GeV were classified as non-showering muons, and those larger than 1~GeV were classified as showering muons.
Criteria for selection of $^8$He candidates were optimized for a good signal to background ratio.
Once a non-showering muon or showering muon was found, the muon was defined as a "current muon". 
According to the end-point energy of the prompt signal~(10.66~MeV, $\beta$ and $\gamma$) and the lifetime~(171.7~ms) of $^8$He $\beta$ decays, the prompt candidate was required to have a reconstructed energy from 4 to 12~MeV in a (30, 500)~ms time window after the current muon.
Considering the decay scheme of $^8$Li, the delayed candidate was required to have energy from 4 to 20~MeV in a (150, 2000)~ms time window after the prompt candidate to retain child $^8$Li and suppress correlated background such as $^{12}$B-$^{12}$B pairs which are discussed below.
The distance between the prompt and delayed candidates was required to be less than 30~cm.
For non-showering muons, the signal to background ratio was enhanced by requiring at least one neutron capture signal in the (10, 300)~$\mu$s window following the muon.
A multiplicity cut required no event with reconstructed energy larger than 0.7~MeV in a $\pm$200~$\mu$s time window about both the prompt and delayed candidates to suppress IBD events.
Events containing additional muons following the current muon are vetoed. The veto time window depended on both the reconstructed muon energy in the AD and the EH to exploit the differing rates and energies of incident muons.

There were 6938, 6409, and 2211 candidates in the three EHs and showering muons contributed 91.9\%, 94.2\%, and 96.5\%, respectively.
Two significant backgrounds were identified.
The first component was due to an accidental coincidence between two events not correlated with the current muon. The second component was correlated background from two isotopes produced by the current muon.
Other background components contributed less than 1\% and were neglected.
A two-dimensional unbinned maximum likelihood fit that considered the spatial and the temporal information was used to determine the signal rate from the two background components.

\begin{figure}[htb]
\centering
    \includegraphics[width=9cm]{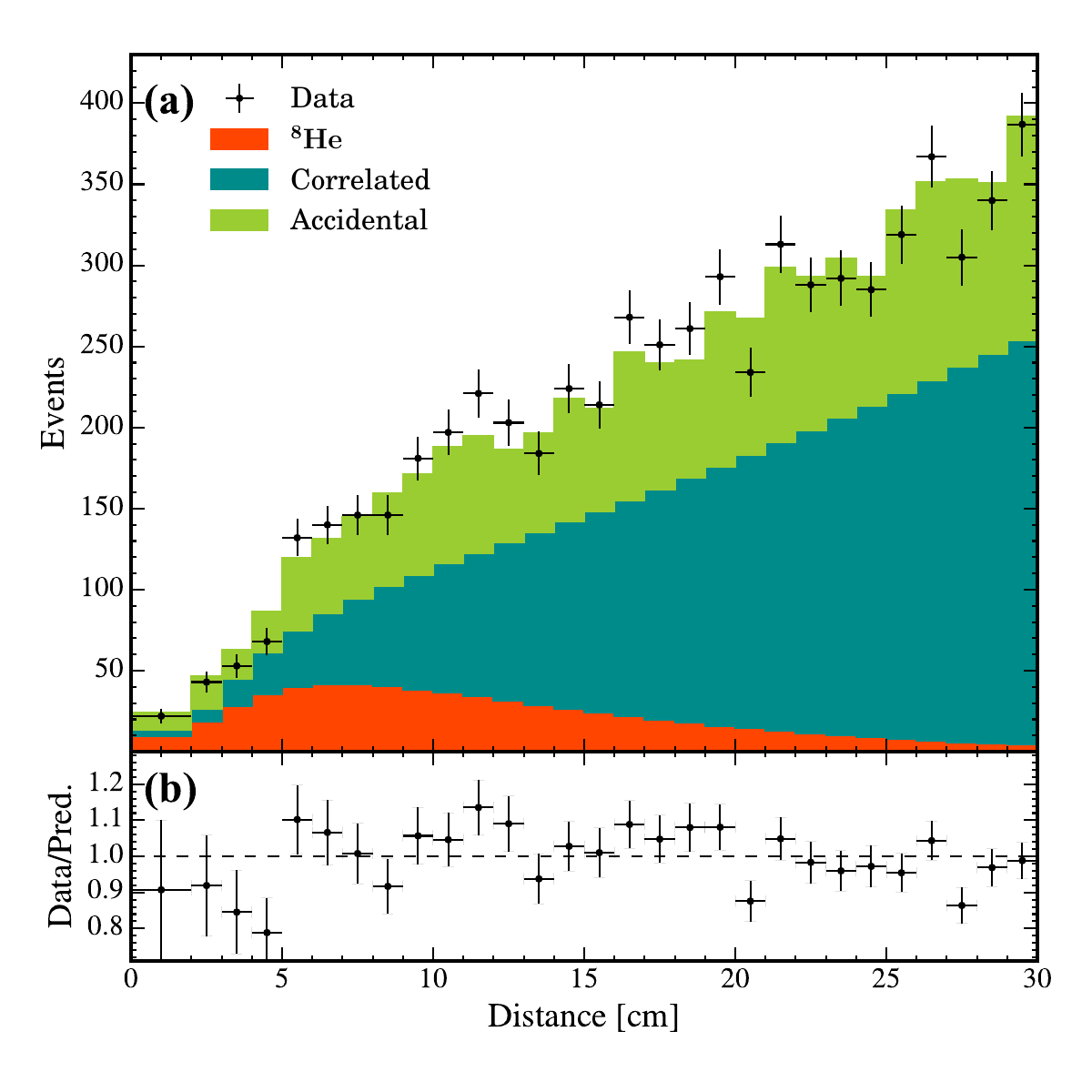}
    \caption{The distribution of distance between the prompt and delayed candidates for showering muons in EH1~(a). Color regions are best-fit predictions and have good agreement with data~(b). The monotonically increasing backgrounds clearly differ from the $^{8}$He signal which peaks at 7~cm. }
    \label{fig:He8Fit}
\end{figure}

The most discriminating variable was the distance~($\Delta R$) between the prompt and the delayed candidates as shown in Fig.~\ref{fig:He8Fit}.
The $\Delta R$ distribution for signal was obtained from simulation.
Because $^8$He and $^8$Li decay at almost the same position, $\Delta R$ featured a narrow distribution peaked at 7~cm primarily due to the resolution of vertex reconstruction.
The $\Delta R$ distribution of accidental coincidences was obtained from the time window before the current muon to ensure that there is no contamination from isotopes produced by the current muon.
The $\Delta R$ distribution of correlated coincidences was from an enriched spallation isotopes sample produced by muons with reconstructed energy $>$2~GeV.
The monotonic increase in both background components as a function of $\Delta R$ provided for powerful discrimination between signal and background.

The second variable was the time interval between the prompt signal and the current muon~($\Delta T$). 
For signals, the $\Delta T$ distribution followed the decay time of $^8$He. 
The prompt signal of the correlated background can be due to one of six primary isotopes: $^{12}$N, $^{12}$B, $^9$C, $^9$Li, $^8$B, and $^8$Li.
Their $\Delta T$ distributions were calculated according to their known lifetimes~\cite{2004nuclei,2017nuclei}.
The accidental coincidences featured a flat $\Delta T$ distribution.

A log-likelihood was defined as 
\begin{align} 
\label{Eq:Fitter8He}
L= \min_{\vec{\bf x}}\left( -2 \sum_{c} \sum_{i} \ln F(\vec{\bf x};\Delta T_{i}, \Delta R_{i},\vec{\bf x} ) + g(\vec{\bf x}) \right),
\end{align}
where $c$ sums over six categories~(showering muons and non-showering muons in three EHs) and $i$ sums over the number of candidates in this category. Free parameters $\vec{\bf r}$ were the ratios of number of events in each type~($^8$He signal, accidental background, and five kinds of correlated background) over the total event number.
Systematical uncertainties $\vec{\bf x}$ were constrained via the pull term $g(\vec{\bf x})$.
The dominant systematic uncertainty was the simulated $\Delta R$ distribution of $^8$He signals.
The shape of $\Delta R$ distributions of naturally occurring $^{214}$Bi and $^{214}$Po decays agreed to within 2\% for simulated and observed events. The discrepancy was parameterized by a linear function in $\Delta R$ which was applied to the $\Delta R$ distribution of $^{8}$He signals in the fit.
Systematic uncertainty in the $\Delta T$ distributions due to the lifetime uncertainties was negligible.
%

Using Eq.~\ref{Eq:Fitter8He}, the numbers of $^8$He events produced by showering muons was determined to be 640.6$\pm$71.1, 751.1$\pm$69.4, and 253.8$\pm$35.7 in the three EHs, respectively.
For non-showering muons, the fitted number of $^8$He was 136.7$\pm$22.9, 109.5$\pm$22.0, and 31.6$\pm$8.2.
About 90\% of the quoted uncertainties are from statistics.
%
The robustness of the best-fit results was examined with ancillary studies. 
Consistent results with 6\% to 8\% larger uncertainties were obtained when only the $\Delta R$ information was used in the fit. 
The measured prompt energy ($E_p$) spectra were consistent with the best-fit prediction. 
Performing a 2-D fit in $\Delta T$ and $E_p$ for the showering muon sample also yielded consistent results.

\begin{figure}[htb]
\centering
    \includegraphics[width=\columnwidth]{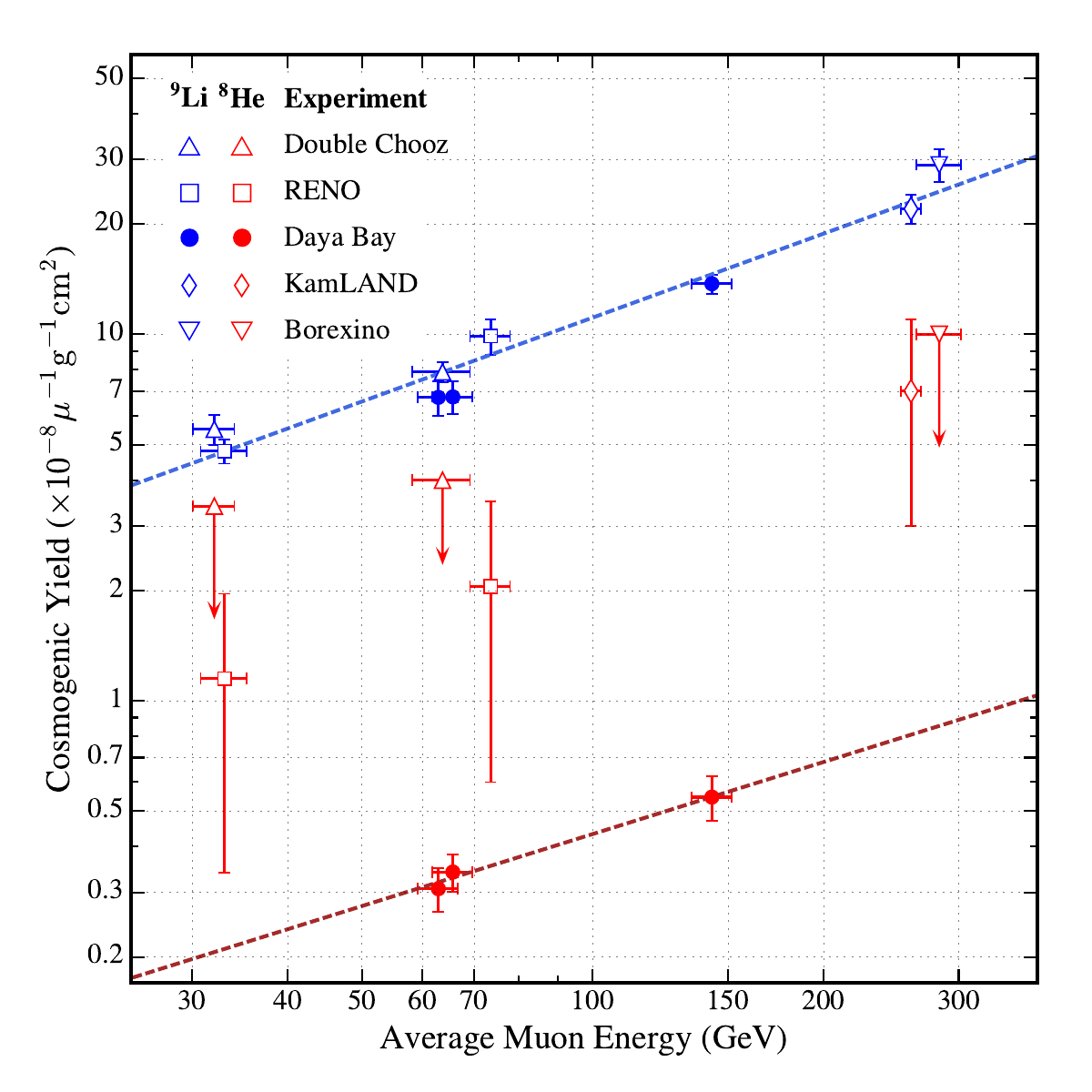}
    \caption{Production yields of $^8$He and $^9$Li by cosmic-ray muons. Experimental results from Double Chooz~\cite{DoubleChooz:2018kvj}, RENO~\cite{RENO:2022xbr}, KamLAND~\cite{KamLAND:2009zwo} and Borexino~\cite{Borexino:2013cke} are also plotted. The points with arrows depict the 2$\sigma$ upper limit. For $^8$He, the dashed line is the power-law fit to the measured results at Daya Bay. For $^9$Li, the fit is performed on all the experimental results. }
    \label{fig:he8yields}
\end{figure}

Requiring at least a neutron capture signal after non-showering muons significantly improved the signal over background ratio and made the fit feasible.
Efficiencies of the neutron tagging were estimated to be (56.0$\pm$20.0)\% , (54.2$\pm$20.0)\% , and (68.6$\pm$25.0)\%  using a control sample of cosmogenic $^{12}$N because neutron tagging efficiency was comparable for cosmogenic production of $^8$He and $^{12}$N in simulation.
The selection efficiency of the energy cuts, the time correlation cuts, and the distance cut was estimated to (33.2$\pm$0.5)\% using the simulation.
Efficiencies of the multiplicity cut and the muon veto were estimated as 59.7\%, 63.2\% and 91.0\% with negligible uncertainties for the three EHs, respectively.
The $^8$He production yields were calculated using the formula:
\begin{align} 
\label{Eq:He8yield}
Y_{\rm He} = N_{\rm He}/(N_{\mu} \times \rho \times {\cal B} \times L^\mu_{\rm LS}),
\end{align}
where $N_{\rm He}$ is the number of $^8$He candidates after efficiency correction, $\rho$ is the density of LS~(0.86~g/cm$^3$) and ${\cal B}$ is the $^8$He $\beta$ decay branching ratio~(Table~\ref{table:he8decay}).
For the three EHs, the yields of $^8$He are 0.307$\pm$0.042, 0.341$\pm$0.040, and 0.546$\pm$0.076 $10^{-8}\mu^{-1}{\rm g}^{-1}{\rm cm}^{2}$ at average muon energies of 63.9~GeV, 64.7~GeV, and 143.0~GeV, respectively~(Fig.~\ref{fig:he8yields}).

This is the first conclusive evidence of $^8$He production by cosmic ray muons in liquid scintillator. 
The yields of $^8$He are approximately 20 times smaller than $^9$Li for muon energies between 60~GeV and 150~GeV.
The $^8$He yield observed at Daya Bay is notably the smallest for any cosmogenic isotope documented so far. 
As illustrated in Fig.~\ref{fig:he8yields}, fitting the $^8$He yields with respect to the average muon energies with the power-law function, $
Y=Y_0\times (\rm E^\mu_{\rm avg}/{\rm 1~GeV})^\alpha$,
gave $Y_0$= $(2.14^{+3.44}_{-1.32})\times10^{-10} \mu^{-1}{\rm g}^{-1}{\rm cm}^{2}$ and $\alpha$= $0.65\pm0.22$.
%

The $^9$Li yields are measured from the same data set using two independent methods that exploit the $\beta$-n channel. Production of $^8$He is neglected in this analysis since the rate in this channel is only about 1.5\%.
The first method was introduced in the Daya Bay neutrino oscillation analysis~\cite{DayaBay:2023orm}.
The selected $^9$Li candidates were paired with the muons within $\pm$2 seconds. 
Both the prompt energy ($E_{\rm p}$) and the time from the nearby muon to the prompt signal ($\Delta T$) were used to separate $^9$Li from IBD events. 

\begin{figure}[h]
\centering
    \includegraphics[width=\columnwidth]{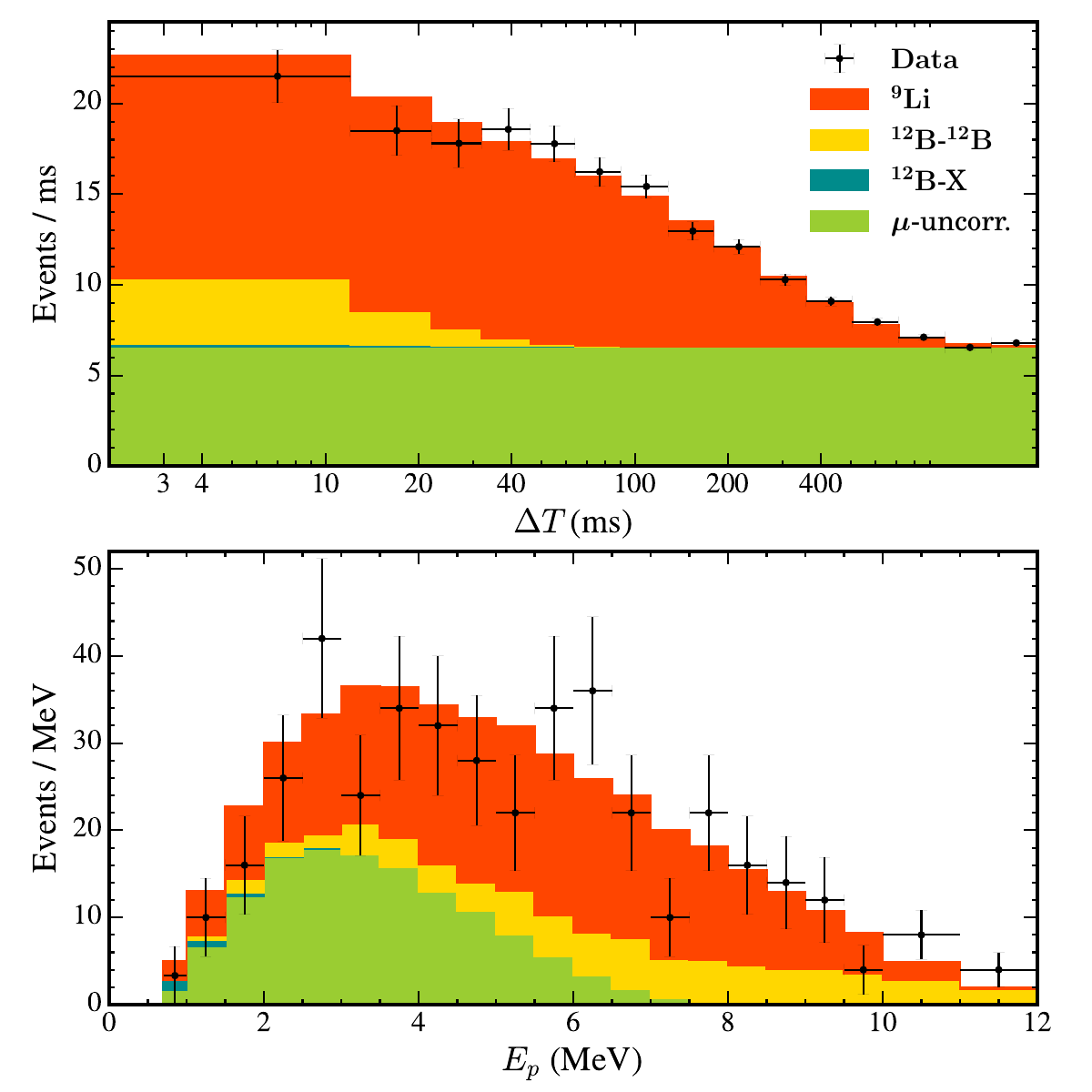}    
    \caption{The $^9$Li and other components are determined by simultaneously fitting 2D distributions including energy ($E_{\text p}$) and time to the nearby muons ($\Delta T$) of the selected IBD prompt signals. For demonstration purpose, the projected 1D distributions of $E_{\text p}$ (with $2$ ms $<\Delta T <12$ ms) and $\Delta T$ (with $0.7$ MeV $< E_{\text p} < 12$ MeV) for the EH3 IBD candidates with $\Delta R<$ 0.5 m are shown.}  
    \label{fig:time}
\end{figure}

The above sample is further divided into 18 sub-samples to improve the sensitivity, according to the experimental halls (EH1, EH2, and EH3), reconstructed muon energies ($E^{rec}_{\mu}<1$ GeV, $1$ GeV $< E^{rec}_{\mu} <2$ GeV and $E^{rec}_{\mu}>2$ GeV), and the distance between  prompt and delayed signals ($\Delta R < 0.5$ m and $\Delta R > 0.5$ m).
%
%
As an example, distributions of $E_{\text p}$ and $\Delta T$ of signal candidates in EH3 for shower muons ($E^{rec}_{\mu}>2$ GeV)
in the $\Delta R < 0.5$ m regions are shown in Fig.~\ref{fig:time}. 

A $\chi^2$ function was defined based on the binned Poisson likelihood ratio method~\cite{Workman:2022ynf} to fit the 18 two-dimensional ($\Delta T$ vs. $E_{\rm p}$) histograms simultaneously. 
Only the $^9$Li decay time and prompt energy spectra were correlated among different sub-samples.
The others were independently determined from the data. 
The goodness of the fit was $\chi^{2}$/NDF = $8892/8985$.
The daily rates of $\beta$-n decays from $^9$Li, $R_{\rm Li}$, were measured as $5.36\pm0.50$, $4.00\pm0.34$, and $0.55\pm0.03$ in three EHs.
The fit results are consistent for different binning strategies of $E^{rec}_{\mu}$ and $\Delta R$.  

%

The second method used the per-event $\Delta T$, $z$ and $w$ information, similar to method used to observe high-energy reactor antineutrinos~\cite{DayaBay:2022eyy}, where $z$ is the vertical position in the AD and $w$ is the weighted reactor power. This method obtained results consistent with the first method.

The $^9$Li yields, $Y_{\rm Li}$, were calculated with the following formula,
$Y_{\rm Li} = R_{\rm Li}/(R_{\mu}\times P_{\mu}\times L^\mu_{\rm GdLS}\times \rho\times {\cal B}\times \epsilon_{\rm Li})$. 
$R_{\mu}$ is the muon rate in one AD with $E^{rec}_{\mu}$ larger than 20~MeV, $21.04$ Hz, $15.59$ Hz and $1.05$ Hz for the three EHs, respectively. 
$P_{\mu}$ is the probability of those AD muons passing through the GdLS region, and $\rho$ is the GdLS density. 
Their values are 62.4\% and 0.86 g/cm$^3$, respectively~\cite{DayaBay:2018heb}. 
%
%
$\cal B$ is the branching ratio of $\beta$-n decay for $^9$Li.
$\epsilon_{\rm Li}$ is the $^9$Li detection efficiency which is estimated to be (80.0$\pm$0.6)\% using simulation~\cite{DayaBay:2018heb}.
Production yields of $^9$Li for the three EHs are $6.73\pm0.73$, $6.75\pm0.70$ and $13.74\pm0.82$ $10^{-8}\mu^{-1}{\rm g}^{-1}{\rm cm}^{2}$.
They are plotted in Fig.~\ref{fig:he8yields} together with results from other LS-based experiments~\cite{DoubleChooz:2018kvj,RENO:2022xbr,KamLAND:2009zwo,Borexino:2013cke}.
Assuming no systematic correlation between experiments, we fit the $^9$Li yields from all experiments with a power-law function and determine $Y_0 = (0.33\pm0.07)\times10^{-8} \mu^{-1}{\rm g}^{-1}{\rm cm}^{2}$ and $\alpha = 0.76\pm0.05$ with a $\chi^2$/NDF of $8.88/7$.

\begin{figure}[htb]
\centering
    \includegraphics[width=\columnwidth]{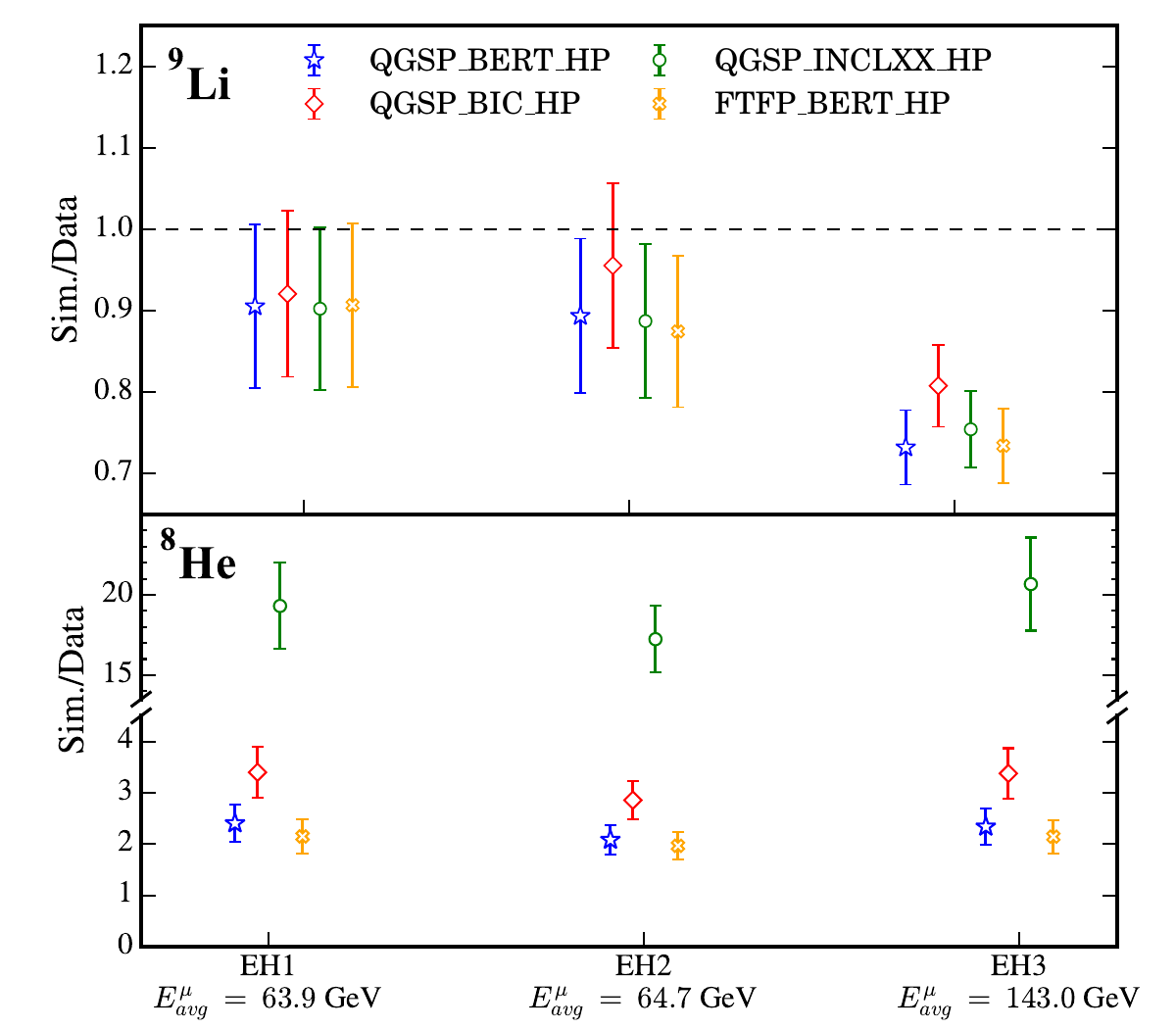}
    \caption{Ratio of simulated results obtained based on different physical models to measured results at Daya Bay. GEANT4 version 11.0.3 was used for simulation. $\rm E^\mu_{\rm avg}$ is the average muon energies at the three experimental halls of Daya Bay.}
    \label{fig:simulation}
\end{figure}

%
The yields are compared with simulated values of GEANT4 version 11.0.3~\cite{Allison:2016lfl} as drawn in Fig.~\ref{fig:simulation}.
Tabulated data can be found in the Supplemental Materials.
Four combinations of hadronic physics lists, QGSP$\_$BERT$\_$HP, QGSP$\_$BIC$\_$HP, QGSP$\_$INCLXX$\_$HP, and FTFP$\_$BERT$\_$HP were used~\cite{G4:Model}.
For $^8$He, the QGSP$\_$INCLXX$\_$HP model gives the worst prediction. 
Predicted values from the other three models are 2 to 4 times larger than the measurement.
For $^9$Li, the simulated values agree between the four models and are 10\%~(25\%) smaller than the measurements in EH1 and EH2~(EH3). 
%

In summary, the individual yields of cosmogenic $^8$He and $^9$Li have been measured by the Daya Bay experiment.
%
This is the first definitive measurement of $^8$He isotope production by cosmic-ray muons in liquid scintillator.
The rate is over ten times smaller than any previously measured cosmogenic isotope production rate.
The measured yields and the power-law relationship with respect to muon energy may help future experiments determine backgrounds to fully exploit their physics potential.
The innovative method to measure the $^8$He yield could be used in other underground liquid-scintillator-based experiments~\cite{2012Chooz,2012RENO,KamLAND:2008dgz,JUNO:2022hxd,Borexino:2013cke,KamLAND-Zen:2023spw}.

\section{Acknowledgement}
The Daya Bay experiment is supported in part by the Ministry of Science and Technology of China, the U.S. Department of Energy, the Chinese Academy of Sciences, the CAS Center for Excellence in Particle Physics, the National Natural Science Foundation of China, the Guangdong provincial government, the Shenzhen municipal government, the China General Nuclear Power Group, the Research Grants Council of the Hong Kong Special Administrative Region of China, the Ministry of Education in Taiwan, the U.S. National Science Foundation, the Ministry of Education, Youth, and Sports of the Czech Republic, the Charles University Research Centre UNCE, the Joint Institute of Nuclear Research in Dubna, Russia, the National Commission of Scientific and Technological Research of Chile, We acknowledge Yellow River Engineering Consulting Co., Ltd., and China Railway 15th Bureau Group Co., Ltd., for building the underground laboratory. We are grateful for the ongoing cooperation from the China Guangdong Nuclear Power Group and China Light \& Power Company.
\bibliographystyle{apsrev4-1}
\bibliography{He8}

\end{document}